\documentstyle[mprocl,epsfig]{article}

\def\be{\begin{equation}}
\def\ee{\end{equation}}
\def\bea{\begin{eqnarray}}
\def\eea{\end{eqnarray}}

\begin{document}

\title{LOW TEMPERATURE SUPERFLUID RESPONSE OF
HIGH-$T_C$ SUPERCONDUCTORS}

\author{T. XIANG$^{1,2}$, C. PANAGOPOULOS$^1$, AND J. R. COOPER$^1$}

\address{$^1$Interdisciplinary Research Center in
Superconductivity, University of Cambridge, Madingley Road,
Cambridge CB3 0HE, United Kingdom}

\address{$^2$Institute of Theoretical Physics, Academia Sinica, 
P.O.Box 2735, Beijing 100080, China}

\maketitle

\begin{abstract}

We have reviewed our theoretical and experimental results of
the low temperature superfluid response function
$\rho_s^\mu$ of high temperature superconductors (HTSC).  In
clean high-$T_c$ materials the in-plane superfluid density
$\rho_s^{ab}$ varies linearly with temperature.  The slope
of this linear $T$ term is found to scale approximately with
$1/T_c$ which, according to the weak coupling BCS theory for
a $d$-wave superconductor, implies that the gap amplitude
scales approximately with $T_c$.  A $T^5$ behavior of the
out-of-plane superfluid density $\rho_s^c$ for clean
tetragonal HTSC was predicted and observed experimentally in
the single layer Hg-compound ${\rm HgBa_2CuO_{4+\delta}}$.
In other tetragonal high-$T_c$ compounds with relatively
high anisotropy, such as ${\rm
Hg_2Ba_2Ca_2Cu_3O_{8+\delta}}$, $\rho_s^c$ varies as $T^2$
due to disorder effects.  In optimally doped ${\rm
YBa_2Cu_3O_{7-\delta}}$, $\rho_s^c$ varies linearly with
temperature at low temperatures, but in underdoped ${\rm
YBa_2Cu_3O_{7-\delta}}$, $\rho_s^c$ varies as $T^2$ at low
temperatures; these results are consistent with our
theoretical calculations.

\end{abstract}

\section{Introduction} 

A growing number of theoretical and experimental works have
showed that the high-$T_c$ superconducting pairing has
$d_{x^2-y^2}$ symmetry \cite{Annett96}.  The existence of
energy gap nodes in this state has profound consequences for
the low temperature electromagnetic response.  In high
quality single crystal or magnetically aligned powder
high-$T_c$ superconductors (HTSC), it has been found that
the in-plane penetration depth $\lambda_{ab}$, whose
temperature ($T$) dependence is determined by the thermal
excitations of unpaired electrons, increases linearly with
$T$ at low $T$, in contrast to the activated behavior of
conventional superconductors.  The intrinsic linear-$T$
behavior of $\lambda_{ab}$ was first observed in a twinned
${\rm YBa_2Cu_3O_{7- \delta}} $ single crystal
\cite{Hardy93} and was one of the earliest pieces of
evidence for the $d$-wave symmetry of the gap parameter.
The linear-$T$ behavior of $\lambda_{ab}$ has now been
established in both chain and nonchain copper oxides, which
include optimally doped or underdoped ${\rm YBa_2Cu_3O_{7-
\delta}} $ \cite{Hardy93,Zhang94,Bonn95,Pana97d,Pana98},
${\rm Bi_2Sr_2Ca Cu_2 O_{8 + \delta} }$
\cite{Jacobs95,Lee96}, ${\rm HgBa_2CuO_{4+\delta} }$
\cite{Pana97a}, ${\rm HgBa_2Ca_2Cu_3O_{8+\delta} }$
\cite{Pana97a,Pana96a}, ${\rm Tl_2Ba_2CuO_{6+\delta} }$
\cite{Broun97}, and ${\rm La_{2-x}Sr_xCuO_4}$
\cite{Pana98b}.

The temperature dependence of the $c$-axis penetration depth
has also been measured in a variety of high-$T_c$ compounds,
such as ${\rm YBa_2Cu_3O_{7- \delta}}$
\cite{Bonn95,Pana97d,Pana97a}, ${\rm La_{2-x}Sr_xCuO_4}$
\cite{Shib94}, ${\rm Bi_2Sr_2Ca Cu_2 O_{8 + \delta} }$
\cite{Jacobs95}, ${\rm HgBa_2CuO_{4+\delta} }$ and ${\rm
HgBa_2Ca_2Cu_3O_{8+\delta} }$ \cite{Pana97a}.  A common
feature revealed in all measurements is that the $T$
dependence of the $c$-axis penetration depth is much weaker
than its in-plane counterpart at low $T$.  It has been found
that $\lambda_c (T)$ approaches to its zero temperature
value as a power law, namely $\lambda_c(T)-\lambda_c(0) \sim
T^n$, with $n$ oftern being sample dependent and generally
greater than 1.

Understanding why the temperature dependence of $\lambda_c$
in high-$T_c$ superconductors is much weaker than
$\lambda_{ab}$ is not a trivial problem.  Straighforward
extension of the 2D $d$-wave model to anisotropic 3D leads
to line nodes on a warped cylindrical Fermi surface.  This
gives a strong linear-$T$ c-axis response in apparent
contradiction to experiment.

The weak $T$ dependence of $\lambda_c$ is actually an
intrinsic feature of HTSC.  As shown by Xiang and Wheatley
\cite{Xiang96a,Xiang96b}, it is due to the interplay between
the $d$-wave symmetry of the superconducting order parameter
and the underlying Cu $3d$-orbital-based electronic
structure of copper oxides.  For high-$T_c$ compounds with
tetragonal symmetry, they predicted \cite{Xiang96b} that in
the clean limit $\lambda_c$ should behave as $T^5$ at low
$T$.  Recently this $T^5$ behavior was observed
experimentally in the single layer Hg-compound by
Panagopoulos et al \cite{Pana97a}.  As in the case of
$\lambda_{ab}$, impurity scattering can have a strong impact
on the $T$ dependence of $\lambda_c$
\cite{Xiang96b,Radtke96,Hirs97}.  In the temperature regime
where impurity scattering is important, $\lambda_c$
generally behaves as $T^2$ or $T^3$, depending on the type
and strength of the scattering potentials
\cite{Xiang96b,Radtke96,Hirs97} .

It is possible that the normal state of HTSC is a non-Fermi
liquid.  However, in the superconducting state, the behavior
of high-$T_c$ materials does not deviate significantly from
that of conventional metallic superconductors except the
different pairing symmetry.  Moreover, angle-resolved
photoemission spectroscopy (ARPES) experiments \cite{Rand97}
show that BCS quasiparticle excitations exist at $T\ll T_c$.
Thus it is generally agreed that the BCS theory is
applicable to HTSC at least at low $T$.

To understand the interlayer dynamics in the superconducting
phase, a key issue which needs to be clarified is whether
the interlayer hopping of superconducting quasiparticles is
coherent or incoherent.  In normal states the $c$-axis mean
free path is short compared with the $c$-axis lattice
constant.  The semiconducting behavior of the out-of-plane
normal state resistivity $\rho_c$ has generally been taken
as evidence that the $c$-axis transport cannot be described
by conventional three dimensional coherent Block transport.
Microwave \cite{Bonn92} and thermal Hall measurements
\cite{Kris95}, however, revealed that the mean free path
increases by more than six orders of magnitude in the
superconducting state.  The superconducting quasiparticle
scattering rate is therefore much lower than the
extrapolated normal state scattering rate.  Thus it is our
belief that excited quasiparticles form coherent Bloch band
along the $c$ axis in the superconducting state at low $T$.
For ${\rm Bi_2Sr_2Ca Cu_2 O_{8 + \delta} }$ or other
extremely anisotropic compounds, the coherent tunneling is
small and the $c$-axis superfluid response might still be
dominated by the impurity assisted hopping \cite{Radtke96}.
However, for other compounds with tetragonal crystal
symmetry and lower anisotropy, we believe that the coherent
band should give a substantial contribution to the $c$-axis
superfluid response function.

This paper arranges as follows.  In Sec.  \ref{sec2} a
general formula for the superfluid response function is
derived from standard linear response theory.  In Sec.
\ref{sec3}, we analyse the electronic structures of copper
oxides and the low $T$ behaviors of the superfluid response
tensor, concentrating particularly on the properties of
$\lambda_c$.  Sec.  \ref{sec4} discusses the effects of
disorder on $\lambda_c$.  In Sec.  \ref{sec5}, we present
experimental results for both $\lambda_{ab}$ and $\lambda_c$
in a variety of high-$T_c$ materials.  A brief introduction
to the technique we used in obtaining the experimental data
is also given in this section.  Finally, Sec.  \ref{sec6}
summarizes the results.

\section{Superfluid response function} \label{sec2}

According to the London equation, the superfluid density is
inversely proportional to the square of the penetration
depth.  From linear response theory \cite{Scal92}, it can
be shown that the superfluid tensor is given by
\begin{eqnarray}  
\rho_s^\mu  & = &  K_\mu  +  \Lambda_\mu
\label{sup_den}
\\
K_\mu & =& {1\over \Omega }\sum_{k\sigma} {\partial^2
\varepsilon_k\over \partial k_\mu^2} \langle
c^\dagger_{k\sigma} c_{k\sigma} \rangle , 
\label{eqkinetic}
\\ 
\Lambda_\mu & =& - {1\over \Omega }\lim_{ k \rightarrow 0 }
\int_0^\beta d\tau \langle J_\mu (k,\tau ) J_\mu (-k, 0)
\rangle ,
\label{eqlambda}
\end{eqnarray} 
where $\langle \,\, \rangle$ denotes a thermal average,
$\varepsilon_k$ is the band energy of electrons, and $J_\mu
(k) = \sum_{q \sigma} (\partial \varepsilon_q/ \partial
q_\mu ) c^\dagger_{q \sigma} c_{q-k\sigma}$ is the current
operator.  $\Omega$ is the volume of the system.  $K_\mu$
and $\Lambda_\mu$ are the diamagnetic and paramagnetic
responses of quasiparticles to an external field,
respectively.  Generally $K_\mu$ is weakly temperature
dependent but $\Lambda_\mu$ is strongly temperature
dependent.  For a spherical energy band, $\varepsilon_k =
k^2/2m^* - \mu$, $K_\mu = n/m^*$ is temperature independent,
where $n$ is the electron concentration.  Above $T_c$, the
contribution from $\Lambda_\mu$ cancels that from $K_\mu$,
leading to a zero superfluid density.  Below $T_c$,
$\Lambda_\mu$ decreases with decreasing temperature.

In the weak coupling BCS theory, Eqs. (\ref{eqkinetic}) and 
(\ref{eqlambda}) can be simplified to
\begin{eqnarray}
K_\mu & =& {1\over \Omega }\sum_{k} {\partial^2
\varepsilon_k \over \partial k_\mu^2} \left( 1 -
{\varepsilon_k \over E_k}\tanh {\beta E_k \over 2} \right) ,
\\
\Lambda_\mu & = & {1\over \Omega}\sum_k 2\left( {\partial
\varepsilon_k\over \partial k_\mu} \right)^2{\partial
f(E_k)\over \partial E_k},
\end{eqnarray}
for a single band system.  Here $E_k =\sqrt{\varepsilon_k^2
+ \Delta_k^2}$ and $\Delta_k$ is the gap order parameter.
For an anisotropic $d$-wave superconductor with
$\varepsilon_k = (1 / 2m_a^*) (k_x^2 + k_y^2) - 2t_\perp
\cos k_z - \mu$, $\Delta_k = \Delta_0 \cos (2\phi )$, $\phi
= \arctan (k_y / k_x)$, and $m_a^* \ll t_\perp^{-1}$, it is
straightforward to show that
\begin{eqnarray}
\rho_s^{a, b} & \approx & {n \over m_a^*} \left( 1 
- {(2 \ln 2) T \over \Delta_0} \right) , 
\label{sab_ani}
\\
\rho_s^c & \approx & {2 m_a^* t_\perp^2 \over \pi} 
\left( 1 - {(2 \ln 2) T \over \Delta_0} \right) ,
\label{sc_ani}
\end{eqnarray} 
up to the leading order approximation of $T$ and $t_\perp$.
$\rho_s^c$ is different than $\rho_s^{a,b}$ because of the
anisotropy, but the normalized superfluid density
$\rho_s^{\mu} (T) / \rho_s^{\mu} (0)$ is approximately the
same in all three directions.  The linear $T$ dependence of
$\rho_s^\mu$ is a direct consequence of the linear low
energy density of states of a $d$-wave superconductor.

The linear $T$ dependence of $\rho_s^{a,b}$ in this
anisotropic system is consistent with experimental results
of HTSC.  However, the $T$ dependence of $\rho_s^c$ given by
Eq.  (\ref{sc_ani}) is too strong compared with experiments,
which indicates that this simple model of an anisotropic
$d$-wave superconductor is insufficient to account for the
experimental data.

\section{Low $T$ behaviors of superfluid tensor} \label{sec3}

In this section, we discuss the properties of the electronic
structure of HTSC and analyse why $\rho_s^c$ has a weaker
$T$ dependence than $\rho_s^{ab}$.  We consider first the
tetragonal (or almost tetragonal) HTSC, and then ${\rm
YBa_2Cu_3O_{7-\delta}}$ materials.  In ${\rm YBa_2Cu_4O_8}$,
where two neighboring CuO chain layers are offset by $b/2$,
the coupling between two neighboring CuO chain layers is
therefore very different from that between a CuO layer and a
CuO$_2$ layer.  To understand the properties of $\rho_s^\mu$
of ${\rm YBa_2Cu_4O_8}$, a proximity $(NSSN)(NSSN)^\prime$
model should be considered, where $S$ represents a CuO$_2$
layer, $N$ a CuO layer, and $(NSSN)^\prime$ represents a
unit cell which is offset by $b/2$ along the $b$-axis with
respect to its neighboring unit cell $(NSSN)$.  At present a
detailed study for this complicated $(NSSN)(NSSN)^\prime$
model has not been made.

\subsection{Tetragonal compounds}

\begin{figure}
\begin{picture}(6, 6)
\put(4,0)
{\epsfig{file = 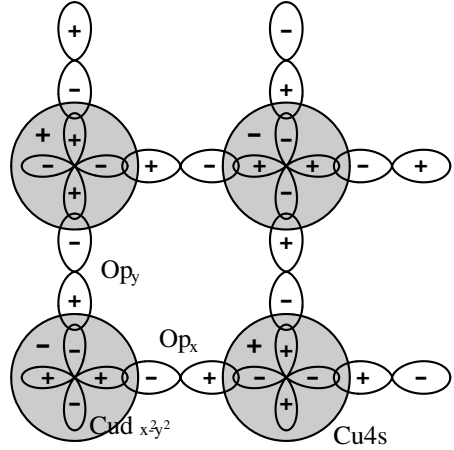, 
width = 6cm, clip = true, 
bbllx = 4.5cm, bblly = 9.5cm, bburx = 19cm, bbury = 24cm } }
\end{picture}
\caption{Cu 3$d_{x^2-y^2}$ and 4$s$ and O 2$p_x$ and 2$p_y$ 
orbitals on each CuO plane.}  
\label{fig1}
\end{figure}

The electronic structure of high-$T_c$ oxides is mainly
determined by the O $2p_x$, $2p_y$ and Cu $3d_{x^2-y^2}$,
$4s$ orbitals (Fig.  \ref{fig1}).  The wavefunctions of the
first three orbitals extend mainly along the short Cu-O bond
axes, leading to a large Cu-O hopping integral in the $ab$
plane.  The O $2p$ can be classified as bonding and
non-bonding Wannier orbitals according to their interaction
with Cu 3$d$ orbitals.  The non-bonding O $2p$ orbital,
which has a zero wavefunction overlap with Cu $3d$, is
believed to be filled and has little effect on the low $T$
transport properties of cuprates.  Thus only the bonding O
orbitals need be considered.  In the limit of strong Coulomb
repulsion, each doped hole on the bonding O $2p$ orbital
will form a local Zhang-Rice spin singlet with a Cu
spin \cite{ZhangRice}.  The low energy physics of high-$T_c$
oxides is governed by this singlet band.  The Cu $4s$ has a
much higher energy than the Fermi level.  However, it is
important for the interlayer hopping of holes
 \cite{Andersen95}, since the interlayer hopping is mainly
assisted by the Cu $4s$ orbitals.  It also has a large
contribution to the 2nd and 3rd nearest neighbor ($t^\prime$
and $t^{\prime\prime}$) intra-plane hopping integrals.

The interlayer hopping of holes is accomplished through the
following virtual hopping processes  \cite{Xiang96b,Xiang97}:
a hole first hops from an O 2p to a Cu 4s orbital on one
CuO$_2$ plane, then hops to another Cu 4s on an adjacent
CuO$_2$ plane via some intermediate orbitals between these
two planes, and finally hops to an O 2p orbital on the
second plane.  Schematically, this interlayer hopping
process can be represented as
$$
\rm (O\,\, 2p)_1\rightarrow (Cu\,\, 4s)_1 
\rightarrow (*)_{12} \rightarrow (Cu\,\, 4s)_2 
\rightarrow (O\,\, 2p)_2, \label{chop}
$$ 
where the subscripts `1' and '2' denote the first and second
CuO layer, respectively.  $(*)_{12}$ represents all relevant
orbitals between two CuO layers.  A doped hole has a certain
probability to occupy the Cu 3$d_{x^2-y^2}$ orbitals.
However, as Cu 3$d_{x^2-y^2}$ and $4s$ orbitals are
orthogonal to each other, a Cu $3d$ hole cannot hop directly
to a Cu $4s$ orbital.

The hopping integral from ${\rm (O\,\, 2p)_1}$ to ${\rm
(O\,\, 2p)_2}$ is proportional to the product of each
individual virtual hopping process, hence the interlayer
hopping integral, $$ \varepsilon_\perp \propto \langle
(2p)_2|(4s)_2 \rangle \langle (4s)_2| (*)_{12}\rangle
\langle (*)_{12} | (4s)_1\rangle \langle (4s)_1 |
(2p)_1\rangle .  $$ The wavefunction overlap between Cu $4s$
and $(*)_{1-2}$ is generally expected to be weakly dependent
on $k_\parallel = (k_x, k_y)$.  However, the wavefunction
overlap between the bonding O $2p$ and Cu $4s$ state has
$d_{x^2-y^2}$ symmetry (Fig 1), i.e.  $\langle (4s) |
(2p)\rangle$ is proportional to $\cos k_x-\cos k_y$, in each
tetragonal CuO$_2$ plane.  Thus $\varepsilon_\perp \propto
(\cos k_x -\cos k_y)^2$, which vanishes along two diagonal
directions of the in-plane Brillouin zone.  This is a unique
property of high-$T_c$ oxides.

The relation $\varepsilon_\perp (k_\parallel ) \propto (\cos
k_x -\cos k_y)^2$ implies that there is no splitting in the
band energy of electrons along the $(0,0)-(\pi , \pi)$
direction for different $k_z$ cross sections.  This property
was actually first found in LDA band structure calculations
\cite{Andersen95} and is supported by all LDA calculations
of high-$T_c$ cuprates with tetragonal symmetry
\cite{Novikov}.

The above discussion indicates that the factor $(\cos k_x
-\cos k_y)^2$ in $\varepsilon_\perp$ comes from the {\it
in-plane} wave function overlap in two CuO$_2$ planes.  Thus
$\varepsilon_\perp\propto (\cos k_x -\cos k_y)^2$ holds for
all high-$T_c$ cuprates with tetragonal symmetry,
independently of the number of CuO$_2$ layers per unit cell
and the way they are coupled together.  In ${\rm
La_{2-x}Sr_xCuO_4}$ or other body centred tetragonal
systems, Cu atoms do not lie collinearly along the $c$-axis.
This will make the $k_\parallel$ dependence of $t_\perp
(k_\parallel )$ in $\varepsilon_\perp (k_\parallel )=t_\perp
(k_\parallel ) (\cos k_x -\cos k_y)^2$ more complicated, but
will not alter the property $\varepsilon_\perp \propto (\cos
k_x -\cos k_y)^2$.

There is a small orthorhombic distortion in ${\rm Bi_2Sr_2Ca
Cu_2 O_{8 + \delta} }$, ${\rm La_{2-x}Sr_xCuO_4}$, and a few
other high-$T_c$ compounds.  However, as the orthorhombic
distortion is along the diagonal direction of the square Cu
lattice on CuO planes in these materials, $\varepsilon_\perp
\propto (\cos k_x -\cos k_y)^2$ is still valid even though
the $x$ and $y$ axes are not exactly perpendicular to each
other.

In high-$T_c$ oxides, Cu $4s$ are always mixed with Cu
$3d_{z^2}$ orbitals.  Thus in the above discussion a Cu $4s$
state should be taken as a mixed state of Cu $4s$ and
$3d_{z^2}$ orbitals.  This mixing, however, does not alter
the above discussion because the Cu 3d$_{z^2}$ orbital, like
Cu 4s, has rotational symmetry about the $c$-axis as well as
reflection symmetry in the CuO plane.  The overlap between
the bonding O $2p$ and Cu 3d$_{z^2}$ orbitals is also
proportional to $(\cos k_x - \cos k_y)$.

Knowing that $\varepsilon_\perp (k) = t_\perp (k_\parallel)
( \cos k_x - \cos k_y )^2$, we can now analyse the
temperature dependence of $\rho_s^c$.  Since at low $T$, the
physical properties of a $ d_{x^2-y^2}$-wave superconductor
are determined by the quasiparticle excitations near the
nodes, we can assume $t_\perp (k_\parallel )= t_\perp$ to be
a $k_\parallel$-independent constant.  The zero of
$\varepsilon_\perp (k)$ is located along the same direction
as the $d_{x^2-y^2}$ gap nodes.  This means that the
paramagnetic current is greatly suppressed along the
$c$-axis at low $T$.  Thus the $c$-axis superfluid density
must behave very differently than in an ordinary anisotropic
system where $\rho_s^c$ varies linearly with $T$ [Eq.
(\ref{sc_ani})].  Indeed, as shown by Xiang and
Wheatley \cite{Xiang96a} and later confirmed experimentally 
by the Cambridge
group \cite{Pana97a}, $\rho_s^c$ varies approximately as
$T^5$ at low $T$
\begin{equation}
\rho_s^c (T) \approx {3\over 4}N(\varepsilon_F ) t_\perp^2 
\left[1 - 450 \left( T\over \Delta_0\right)^5 \right],
\label{T_five}
\end{equation}
where $\Delta_0$ is the gap maximum at $T=0$ and
$N(\varepsilon_F)$ is the normal density of states on
CuO$_2$ planes.  This $T^5$ term is from the contribution of
the paramagnetic term:  one power of $T$ is from the linear
density of states of the $d$-wave superconductor, $\rho (E )
\sim E$; the other $T^4$ is from the $(\cos k_x- \cos
k_y)^4$ factor in $(\partial \varepsilon_k/ \partial k_z)^2
\propto \varepsilon_\perp^2(k_\parallel ) $ which is
approximately proportional to $ E^4$ at low energy.  This
$T^5$ power law of $\rho_s^c$ is true only if the high-$T_c$
pairing has the $d_{x^2-y^2}$ symmetry.  If the gap nodes
are not located along the zone diagonal, such as in a
$d_{xy}$-wave superconductor, the low temperature behavior
of $\rho_s^c$ will be completely different.  Therefore, from
the measurement of $\rho_s^c$ we can determine unambiguously
whether the high-$T_c$ pairing has the $d_{x^2-y^2}$
symmetry.

There are several effects which may lead to a finite hopping
integral along the $c$-axis in the vicinity of the gap nodes
in clean systems and thus a stronger $T$ dependence of
$\rho_s^c$.  Direct interplane hoppings which are not
assisted by Cu 4$s$ orbitals, for example, can make the
$c$-axis dispersion finite around the gap nodes and
eliminate the zeros in $\varepsilon_\perp (k_{\parallel})$
altogether.  In this case, we may model the $c$-axis
electronic structure by $\varepsilon_\perp(k_\parallel) =
t_\perp (\cos k_x - \cos k_y )^2 + t_\perp^{\rm node}$,
where $t_\perp^{\rm node}$ denotes the interlayer hopping
integral at the gap nodes contributed from all possible
interlayer hopping channels not assisted by Cu 4$s$ states.
$t_\perp^{\rm node}$ is expected to be small in high-$T_c$
compounds, but it generates a small linear term in
$\rho_s^c$, with a slope proportional to $(t^{\rm
node}_\perp )^2$, which dominates the $c$-axis superfluid
response when $T \ll t_\perp^{node}$.

\subsection{${\rm YBa_2Cu_3O_{7-\delta}}$ materials}

YBCO contains CuO chains.  This is the main feature which
distinguishes YBCO from other high-$T_c$ compounds.
Recently Hardy and co-workers at UBC measured the
penetration depth of untwinned single crystals of
$YBa_2Cu_3O_{7-\delta}$ and $YBa_2Cu_4O_8$ along the three
principal axes \cite{Zhang94,Bonn95,Basov95}.  They found
that the ratio of the superfluid density in $b$ and $a$
directions at zero temperature is $\approx 2.4$ for the
one-chain compound
$YBa_2Cu_3O_{6.95}$ \cite{Basov95,Tallon95} and $\approx 6$
for the two-chain compound $YBa_2Cu_4O_8$ \cite{Basov95}.
Anisotropies of similar magnitudes are observed in the
normal state resistivity as well \cite{Frie90}.  This
indicates that more than half of the $b$-axis superfluid
density comes from the contribution of CuO chains.  Knight shift and
NMR relaxation rate measurements \cite{Imai88,Taki89} in
$YBa_2Cu_3O_{6.95}$ have also revealed that there is an
appreciable gap on the chains below $T_c$.  Thus to
understand the properties of the c-axis electromagnetic
response functions in YBCO, we need consider not only the
properties of electrons on the CuO$_2$ planes, but also
those on the CuO chains.

An interesting feature revealed by the penetration depth
measurement of the one-chain compound ${\rm
YBa_2Cu_3O_{7-\delta}}$ is that the temperature dependences
of the superfluid densities in $a$ and $b$ directions are
similar:  both are linear at low
temperature \cite{Zhang94,Bonn95}, and roughly obey
$\rho_a^{(s)}(T) / \rho_a^{(s)}(0) \simeq \rho_b^{(s)}(T) /
\rho_b^{(s)}(0)$ up to $T_c$.  The linear $T$ dependence of
$\rho_c^{a,b}$ can be understood from the standard theory of
$d$-wave superconductors.  However, the relation
$\rho_a^{(s)}(T) / \rho_a^{(s)}(0) \simeq \rho_b^{(s)}(T) /
\rho_b^{(s)}(0)$ is unexpected.  This unexpected result
indicates that the CuO chain layers must be intrinsically
superconducting, yet there must be a node of the energy gap
on the chain Fermi surface sheet.  Otherwise, one should
observe ${({\rm d} \rho_b^{(s)} / {\rm d}
T)}|_{T=0} \simeq {({\rm d} \rho_a^{(s)}/ {\rm d}
T)}|_{T=0}$, instead of ${\rm d} [\rho_a^{(s)}(T) /
\rho_a^{(s)}(0)] / {\rm d}T |_{T=0} \simeq {\rm d}
[\rho_b^{(s)}(T) / \rho_b^{(s)}(0)] / {\rm d}T |_{T=0}$,
since the chain band gives no contribution to the linear
temperature term of $\rho_b^{(s)}$ at low $T$.  As
a 1D pairing state has a finite energy gap under ordinary
circumstances, the presence of nodes suggests that the gap
function of the chain band must have 2D character, i.e.
inter-chain pairing exists even though direct inter-chain
hopping does not exist in the chain layer.

Because of the 1D character of the CuO chains, the
wavefunction overlap between O $2p$ and Cu $4s$ orbitals on
the CuO chain layers does not have the $d_{x^2-y^2}$
symmetry.  This changes the $k_\parallel$ dependence of
$\varepsilon_{\perp k}$ and leads to a finite
$\varepsilon_\perp$ along $k_x=\pm k_y$.  Besides, the
CuO$_2$ planes in YBCO are dimpled with relative displacements of O
in the $c$ direction.  The O displacements further reduce
the crystal symmetry and introduce a noticeable
hybridization between the $\sigma$ and $\pi$ bands.  This
hybridization will also make $\varepsilon_\perp $ finite
along the zone diagonals.  Thus in YBCO, $\varepsilon_\perp
$ is not simply proportional to $(\cos k_x - \cos k_y)^2$
and $\rho_s^c$ will not follow that characteristic $T^5$
behavior at low $T$.

In order to understand the low temperature behavior of ${\rm
YBa_2Cu_3O_{7-\delta} }$, let us consider a weak coupling
BCS model of alternately stacking planar ``CuO$_2$" and
chain ``CuO" layers along the c-axis
\cite{Xiang96a,Klemm95,Atki96}:
\begin{eqnarray}
H & = & H_0 + H_1 , 
\label{ham_ybco}
\\
H_0 & =& \sum_{k\sigma} \left( \begin{array}{cc}
 c^\dagger_{1k\sigma} & c^\dagger_{2k \sigma} \end{array}
 \right) \left( \begin{array}{cc} \varepsilon_{1k} &
 \varepsilon_{\perp k} \\ \varepsilon_{\perp k} &
 \varepsilon_{2k} \end{array} \right) \left(
 \begin{array}{c} c_{1k\sigma}\\ c_{2k \sigma} \end{array}
 \right),
\nonumber \\
H_1& = & \sum_{\alpha =1, 2 ; k} \Delta_{\alpha k} \left(
 c_{\alpha -k\downarrow} c_{\alpha k \uparrow} + h.c.
 \right) ,
\nonumber
\end{eqnarray}
where $c_{1k\sigma}$ and $c_{2 k \sigma}$ are electron
operators in the plane and chain bands, respectively.  To
mimic the electronic structures of YBCO-type materials, we
assume the energy dispersions of the uncoupled plane and
chain bands to be
\begin{eqnarray}
\varepsilon_{1k} & = & - 2t(\cos k_x+\cos 
k_y) - 4 t^\prime \cos k_x \cos k_y-\mu ,
\nonumber \\
\varepsilon_{2k} & = & -2 t_c \cos k_y -\mu_c,
\nonumber
\end{eqnarray}
where $\mu$ and $\mu_c$ are the chemical potentials of the
plane and chain bands, respectively.  The coupling between
the chains and planes is introduced by the nearest neighbor
interlayer hopping of electrons
$$
\varepsilon_{\perp \, k} = -2 t_\perp \cos(k_z/2), 
$$
which vanishes at the zone boundary $k_c = \pi$.  $t_\perp$
is the interlayer hopping constant and is in general a
function of $k_\parallel$.  However, since $t_\perp$ is
finite in the vicinity of $d_{x^2-y^2}$ gap nodes, we can
take $t_\perp$ as a $k_\parallel$-independent constant, when
we discuss low temperature properties of $H$.  For YBCO,
$t_c$ is of the same order as $t$, but $t_\perp$ is much
smaller than $t$ or $t_c$.

\begin{figure}
\begin{center}
\leavevmode\epsfxsize=8cm
\epsfbox{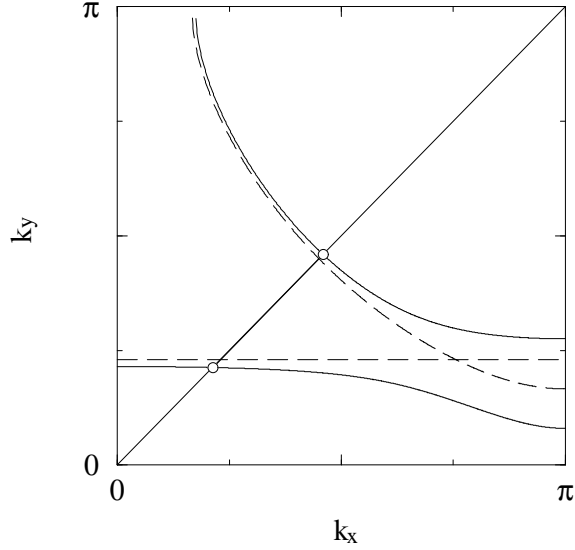}
\caption{Fermi surfaces of ${\tilde \varepsilon}_{1k}$ and 
${\tilde \varepsilon}_{2k}$
bands with $k_z =0$ (solid curves) and $k_z = \pi /c$
(dashed curves).  The circles are $d_{x^2-y^2}$-wave gap
nodes at $k_z =0$ plane.  The parameters used are $t_1 =
-t/4$, $t_c=1.2t$, $t_z = 0.2t$, $\mu = -0.6t$, and $\mu_c =
-1.8t$.
}
\label{fermi_surface}
\end{center}
\end{figure}

The interlayer hopping hybridizes the plane and chain bands, 
and the energy dispersions of the hybridized plane and chain
bands are given by
$$
{\tilde \varepsilon}_{\alpha k } = {\varepsilon_{1k} +
\varepsilon_{2k} \pm \sqrt{ (\varepsilon_{1k} -
\varepsilon_{2k} )^2 + 4 \varepsilon_{\perp k}^2 } \over 2
}, \quad (\alpha =1, \, 2).
$$
When $\varepsilon_{\perp k} \not= 0$, the chain band
acquires dispersion in the $a$ direction.  The Fermi surface
of these hybridized bands is determined by the solution of
$\varepsilon_{1k} \varepsilon_{2k} = \varepsilon^2_{\perp
k}$.  In Fig.  \ref{fermi_surface}, we show the Fermi
surface contours of ${\tilde \varepsilon}_{1 k}$ and
${\tilde \varepsilon}_{2 k}$ on the $k_z=0$ (solid curves)
and $k_z = \pi$ (dashed curves) planes.  The area between
the solid and dashed contours gives the $z$ dispersions of
${\tilde \varepsilon}_{1 k}$ and ${\tilde \varepsilon}_{2
k}$.  The parameters used in Fig.  \ref{fermi_surface} are
chosen so that the Fermi surface contours of ${\tilde
\varepsilon}_{1k}$ and ${\tilde \varepsilon}_{2k}$ resemble
qualitatively those of YBCO as obtained by ARPES
measurements.

In $H_1$, $\Delta_{1 k}$ and $\Delta_{2 k}$ are the gap
order parameters on the plane and chain layers,
respectively.  If pairing potentials exist independently (or
nearly independently) on the planar and chain layers,
$\Delta_{1 k}$ and $\Delta_{2 k}$ are then determined by
self-consistent gap equations 
\begin{equation} 
{1 \over
\Omega}\sum_k V^{(\alpha)}_{k,k^\prime} \langle c_{\alpha
k\uparrow} c_{\alpha \,-k\downarrow}\rangle = \Delta_{\alpha
k^\prime}, \label{gap_eq} 
\end{equation} 
where $V^{(\alpha)}_{k,k^\prime}$ ($\alpha=1$, 2) are
pairing potentials.

When the plane and chain bands are uncoupled, there are two
critical transition $T$, corresponding to the
superconducting transitions of CuO$_2$ planes and CuO
chains, respectively.  The quantum fluctuation of electrons
in the chain band is very strong since CuO chains are
quasi-one dimensional, and the transition temperature of CuO
chains, $T_c^{\rm chain}$, is in general very small compared
with the transition temperature of CuO$_2$ planes, $T_c^{\rm
plane}$.  In this case, $\rho_s^b (T)$ will rise abruptly
below $T_c^{\rm chain}$ where the contribution from the CuO
chains to $\rho_s^b (T)$ becomes finite, as shown in Fig.  2
of Ref.   \cite{Xiang96a}.  Switching on the interlayer
coupling leads to single superconducting phase transition
with $T_c \approx T_c^{\rm plane}$ and a smooth curve of
$\rho_s^b(T)$ in the whole temperature range.  However, as
the energy scale associated with $T_c^{\rm chain}$ still
exists in this system, $\rho_s^b (T)$ will turn
upwards \cite{Xiang96a}, i.e.  have positive curvature
$\partial^2 \rho_s^b (T) / \partial T^2 > 0$, in the
vicinity of $T_c^{\rm chain}$.  For the same reason,
$\rho_s^c (T)$ will also turn upwards, i.e.  $\partial^2
\rho_s^c (T) / \partial T^2 > 0$, in the vicinity of
$T_c^{\rm chain}$.  If the CuO chain layers are not
intrinsically superconducting, namely $V^{(2)}_{k, k^\prime}
=0$ or $T_c^{\rm chain}=0$, it can be further shown that the
leading temperature dependence of $\rho_s^b$ is
\begin{equation}
 \rho_b \sim \sqrt{T}
\end{equation} 
at low $T$. $T$.  This $\sqrt{T}$ behavior of $\rho_s^b$ has
actually been found recently in double-chain ${\rm
YBa_2Cu_4O_8}$ compound \cite{Pana98c}.  However, in ${\rm
YBa_2Cu_3O_{7-\delta}}$, this $\sqrt{T}$ behavior or more
generally the upturn curvature of $\rho_s^{b,c}(T)$, has not
been observed in the penetration depth measurements by 
our group \cite{Pana97d}, neither the UBC group
\cite{Zhang94,Bonn95}.

The absence of a positive curvature in $\rho_s^{b,c}$ of
${\rm YBa_2Cu_3O_{7-\delta}}$ implies that $\Delta_{2k}$
in ${\rm YBa_2Cu_3O_{7-\delta}}$ should have the same order
and follow approximately the same $T$ dependence as
$\Delta_{1k}$.  Based on this analysis, Xiang and Wheatley
 \cite{Xiang96a,jmw} proposed a pairing tunneling model to
account for the experimental data.  In this model, the gap
order parameter of the planar and chain bands are strongly
coupled together and $\Delta_{\alpha k}$ are determined by
\begin{equation}
{1  \over \Omega}\sum_k V_{k,k^\prime} \langle
c_{\alpha k\uparrow} c_{\alpha \,-k\downarrow}\rangle = 
\Delta_{{\bar\alpha}  k^\prime},
\label{pair_tunnel}
\end{equation}
where ${\bar\alpha} = 1$ or 2 if $\alpha = 2$ or 1.
$V_{k,k^\prime}$ is the interlayer pair hopping amplitude.
In Eq.  (\ref{pair_tunnel}) the gap function of the planar
(chain) band is determined by the pair correlation function
of the chain (planar) band.  This is the most important
feature of the pair tunneling model.  It will force
$\Delta_{1k}$ and $\Delta_{2k}$ to have not only the same
order of magnitude with similar temperature dependence but
also the same symmetry to the leading order approximation.
Thus in this model there is only one energy scale and
$\rho_s^{b,c}(T)$ will not turn upwards.

The formulae for the $c$-axis superfluid density in this
plane-chain band model are still given by Eqs.
(\ref{sup_den}) and (\ref{eqlambda}), but the definitions
for $K_c$ and $J_c$ should be changed to
\begin{eqnarray}
K_c & = & {1\over \Omega}\sum_{k\sigma} {\partial^2
\varepsilon_{\perp k} \over \partial k_z^2} \langle
c_{1k\sigma}^\dagger c_{2k\sigma} + h.c.  \rangle,
\\ 
J_c & = & \sum_{k\sigma} {\partial \varepsilon_{\perp k}
\over \partial k_z} ( c_{1k\sigma}^\dagger c_{2k\sigma} +
h.c.  ) .
\end{eqnarray}
To analyse the low temperature behavior of $\rho_s^c$, let
us first consider the case where $\Delta_{1k} = \Delta_{2k}
= \Delta_{k}$ and $\Delta_{k}$ has $d_{x^2-y^2}$
symmetry.  In this case, the energy spectrum of
superconducting quasiparticles is largely simplified, and the
diamagnetic and paramagnetic terms of $\rho_s^c$ are given 
by
\begin{eqnarray}
K_c & = & - {1\over \Omega} \sum_{k\alpha=1,2} {\partial^2
\varepsilon_{\perp } \over \partial k_z^2} {\partial E_\alpha 
\over \partial \varepsilon_\perp } \tanh {\beta E_\alpha \over 2} ,
\\
\Lambda_c & = & {1\over \Omega} \sum_{k} \left[ 
2\sum_{\alpha=1,2} \left( {2
\varepsilon_\perp \over {\tilde \varepsilon}_1 - {\tilde
\varepsilon}_2 } {\partial \varepsilon_\perp \over \partial
k_z } \right)^2 {\partial f(E_\alpha ) \over \partial
E_\alpha } \right. 
\nonumber \\
& & \left. - {A_{k,-} \over E_1 + E_2}
+  \sum_{\sigma =\pm} A_{k,\sigma} {f(E_1) - \sigma
f(E_2) \over E_1 - \sigma E_2} \right] , 
\end{eqnarray}
where 
$$
A_{k,\pm} = 2 \left( {\partial \varepsilon_{\perp } \over
 \partial k_z } \right)^2 { (\varepsilon_1 - \varepsilon_2
 )^2 \over ({\tilde \varepsilon}_1 - {\tilde
 \varepsilon}_2)^2} \left( 1 \pm {{\tilde \varepsilon}_1
 {\tilde \varepsilon}_2 + \Delta^2_k \over E_1 E_2} \right),
$$  
and $E_{\alpha k} = \sqrt{ {\tilde \varepsilon}_{\alpha k}^2
+ \Delta_k^2 } $ is the energy dispersion of quasiparticles
of the $\alpha$ band.  The gap nodes of $E_{\alpha k}$ are
located along the zone diagonal on the Fermi surface of
${\tilde \varepsilon}_{\alpha k}$, as shown in Fig.
\ref{fermi_surface}.

The diamagnetic contribution to $\rho_s^c$, $K_c$, is
proportional to the kinetic energy of the system along the
$c$-axis.  To the leading order approximation of $t_z$, we
find that
\begin{equation}
K_c \approx c_1 {t_z^2 \over 2 t}, 
\label{diamag} 
\end{equation}
which is temperature independent.  The correction to Eq. 
(\ref{diamag}) is of the order of $\exp ( - W / T)$, with $W$
the band width of $\varepsilon_{1k}$ or $\varepsilon_{2k}$,
which is negligible at low $T$.  Here we have
assumed that the energy difference between ${\tilde
\varepsilon}_{+, k}$ and ${\tilde \varepsilon}_{-, k}$ at
the same $k$ point is on average much smaller than $W$.
$c_1$ is a numerical factor which depends on the detailed 
structures of the planar and chain bands.  Numerically we
found that $c_1$ varies between 0.7 and 1.3 in a broad
regime of parameters which are physically relevant.  Thus
approximately we can take $c_1 = 1$.

The first term of $\Lambda_z$ is the intra-band
contribution.  Since the derivative of the Fermi function
${\partial f(E_\alpha) /\partial E_1}$ is nonzero only in
the vicinity of gap nodes of the $\alpha$ band at low
$T$, this term can be readily evaluated.  For the
convenience of the discussion below, we assume that the gap
nodes of the $\alpha$ band are located at $k=Q_\alpha$ in
the Brillouin zone.  Around the gap nodes of the first
(second) band, $({\tilde \varepsilon}_{1k} - {\tilde
\varepsilon}_{2k})^2$ can be approximated by $\varepsilon_{2
Q_1}^2$ (or $\varepsilon_{1 Q_2}^2$) and taken as a constant
out of the integration (or summation).  To the leading 
order approximation in $T$, we find that the first term of 
$\Lambda_c$ is 
\begin{equation}
\Lambda_c^{(1)} \approx - {t_\perp^4  \over 2} 
\left[ {N(\varepsilon_{1,F} )\over \varepsilon^2_{2, Q_1} } + 
{N(\varepsilon_{2,F} )
\over \varepsilon^2_{1, Q_2} } \right] {(2 \ln2) T \over \Delta_0} , 
\end{equation}
where $N(\varepsilon_{\alpha,F})$ is the density of states of the
$\alpha$ band at the Fermi level.  Since
$\varepsilon_{\alpha Q_{\bar \alpha}}$ ($\alpha =1$, 2) is
of order $t$ and $D_{\alpha ,F}$ is of order $t^{-1}$, we
thus have $\Lambda_c^{(1)} \sim t_\perp^4 T / t^3 \Delta_0 $,
which is by a factor of $(t_\perp/t)^2$ smaller than $K_c$.

The second and third terms of $\Lambda_c$ are the inter-band
contributions.  The second term, which is $T$ independent
and of the same order as $K_c$, will cancel part of the
contribution of $K_c$ to $\rho_s^c$.  The third term is
temperature dependent.  To the leading order approximation
in $T$, we find that this term behaves as $T^3$ at low
temperature:
$$
\Lambda_c^{(3)}  \sim { t^2_\perp T^3 \over t^3 \Delta_0 }.
$$
When $T\ll t_\perp$, this term can be ignored since it is
much smaller than $\Lambda_{(1)}$.

We note that in this plane-chain band model the paramagnetic
contribution to $\rho_s^c$ is non-zero even at $T=0$.  This
is a peculiar property of the model.  We believe that this
`residual' paramagnetic current will affect the spin-lattice
relaxation rate and many other properties of the system.  A
more comprehensive study for this model is therefore
required to clarify to what extend it can be used to
describe the physical properties of ${\rm
YBa_2Cu_3O_{7-\delta}}$ materials.

Thus at low $T$
\begin{equation}
\rho_s^c(T) \approx c_1^\prime {t_\perp^2 \over 2t } \left[
1 - c_2 \left( {t_\perp \over t} \right)^2 {(2 \ln 2)T 
\over \Delta_0} + o(T^3) \right] ,
\label{ybco_rho}
\end{equation}
where $c_1^\prime < c_1$ and $c_2$ are band-structure
dependent and dimensionless constants of order 1.  At zero
temperature, the in-plane superfluid density
$\rho_s^{a,b}(T)$ is of order $t$ and $\rho_s^{c}(0) /
\rho_s^{a,b}(0) \sim (t_\perp / t)^2$.  Since the
penetration depth $\rho_\mu \propto 1 / \sqrt{\rho_s^\mu }$,
we have $\lambda_c (0) / \lambda_{a, b} (0) \sim t /
t_\perp$.  Thus from the experimental results of $\lambda_c
(0) / \lambda_{a, b} (0) $ we can estimate the value of
$t_\perp / t$.

In the above discussion, we have assumed that $\Delta_{1k} =
\Delta_{2k}$.  In general, however, $\Delta_{1k}$ is
different from $\Delta_{2k}$, although they may have the
same symmetry and follow approximately the same temperature
dependence, such as in the pair tunneling
model \cite{Xiang96a}.  When $\Delta_{1k} \not= \Delta_{2k}$,
an analytic treatment for the temperature dependence of
$\rho_s^c$ becomes difficult, and we therefore studied the
low temperature behavior of $\rho_s^c$ numerically.  We
found that in addition to the linear $T$ contribution a
$T^2$-term appears in $\rho_s^c$.  In the temperature regime
so far measured, this $T^2$ term is found to be even larger
than the linear $T$ term in a broad regime of parameters
which are physically relevant.  Thus at low $T$, we
should include a $T^2$ term in Eq.  (\ref{ybco_rho}) and
take the coefficient of this $T^2$ term as a fitting
parameter when it is used for comparison with experiment.

The linear $T$ term of the renormalized $c$-axis superfluid
density $\rho_s^c (T) / \rho_s^c(0)$ is by a factor of
$(t_\perp / t)^2$ smaller than the corresponding in-plane
value.  This implies that this linear $T$ term is very
difficult to detect experimentally.  For optimal doped
YBCO, $t_\perp / t \sim 0.2$, it may still be possible to
observe experimentally the contribution of this linear $T$
term.  For underdoped YBCO with $t_\perp / t < 1/20$,
however, the $T^2$ term may dominate the low $T$ behavior
of $\rho_s^c$.

In real YBCO materials, the system cannot be in a pure
$d_{x^2-y^2}$ pairing state because of the strong in-plane
anisotropy.  A more complete self-consistent solution of the
gap equations should allow admixture of an s-wave component
to form a d+s wave state.  Admixture of a small s-component
is consistent with the c-axis Josephson tunneling
experiments on $YBa_2Cu_3O_{7-\delta}$ \cite{Sun94}.  The gap
nodes survive for weak admixture, only the positions of the
nodes shift away from the diagonals of the Brillouin zone.
Thus, qualitatively, the above discussion for $\rho_s^c$ is
still valid.

\section{Disorder effects} \label{sec4}

Disorder effect is an important but very complicated problem
in the analysis of experimental data of HTSC.  For
tetragonal materials, disorder is always a relevant
perturbation since the $c$-axis hopping rate of electrons
always becomes small in comparison with the impurity
scattering rate when the quasiparticle momentum is
sufficiently close to the nodal line.  For YBCO related
materials, the strong localization effect in the CuO chain
layers may have a strong impact on the interlayer coupling.

In a disordered system, the quasiparticle lifetime becomes
finite.  This introduces a finite density of states at the
Fermi level and changes the behavior of the superfluid
tensor.  At low $T$, $\rho_s^{ab}$ should now behave as
$T^2$ since the low energy density of states of
quasiparticles varies quadratically with energy in a
disordered $d$-wave superconductor.  Along the $c$-axis, the
singularity in $\varepsilon_\perp$ becomes less important,
and when $T \ll \Gamma$, $\rho_s^c$ can be calculated as in 
a disordered $d$-wave pairing state with a constant
hopping constant, hence $\rho_s^c$ should also behave as
$T^2$ at low $T$.

Another important aspect of disorder is that it may disrupt
the symmetry of bonding orbitals about the copper site.  The
selection rule preventing hopping for momentum along
$k_x=\pm k_y$ in tetragonal HTSC is then relaxed.  For
example, interlayer defects on O sites may introduce a random
component to $\varepsilon_\perp$.  Near the gap nodes this
fluctuating component in $\varepsilon_\perp$ dominates the
$c$-axis hopping, and thus electrons in the vicinity of
nodes are well described by an impurity assisted hopping
model \cite{Radtke96}.  Impurity assisted hopping gives a new
conduction channel and has a direct contribution to the
$c$-axis superfluid density $\rho_{s,imp}^c$.

The discussion given in this section is limited to the
tetragonal high-$T_c$ materials only.  Both the self-energy
(or lifetime) effect and impurity assisted hopping are
considered.  For YBCO, a systematic analysis of the effects of 
disorder, considering the peculiar properties of CuO
chains, is still not available.

\subsection{Self-energy effect}

This is a commonly studied disorder effect in a $d$-wave
superconductor.  It is known that the correction from the
self-energy of electrons to the diamagnetic response
function $K_\mu$ is small.  Hence $K_{ab}$ and $K_c$ are
still approximately given by the $T$-independent terms in
Eqs.  (\ref{sab_ani}) and (\ref{T_five}), respectively.
However, at low $T$, the correction from the
self-energy of quasiparticles to the paramagnetic response
function $\Lambda_\mu$ is quite significant.  If we ignore
the vertex correction, then $\Lambda_\mu$ is approximately
given by
\begin{equation}
 \Lambda_\mu \approx  -{1\over \pi \Omega} 
\sum_k \left( 
 {\partial \varepsilon_k \over \partial k_\mu } \right)^2
\int  {\rm d} \omega  f(\omega ){\rm Im} {\rm Tr} G^2(k, \omega ),  
\end{equation}
where $G(k, \omega )$ is the retarded electron Green's function
\begin{equation}
G(k, \omega ) = {1\over  \omega - \varepsilon_k \tau_3
+\Delta_k \tau_1 + i  \Gamma }
\end{equation}
and $\tau_{1,3}$ are Pauli matrices.  The low temperature
dependence of $\Lambda_\mu$ can be readily obtained with the
Sommerfeld expansion.  We find that when $T \ll \Gamma \ll
\Delta_0$,
\begin{eqnarray}
\Lambda_{ab}(T)  &\approx &  -{ v_F^2 N(\varepsilon_F) \Gamma 
\over \pi \Delta_0} \left( \ln {\Delta_0\over \Gamma} + 
{\pi^2 T^2 \over 6 \Gamma^2} \right),
\label{ab_time}\\
\Lambda_c(T) &\approx & - {8 t_\perp^2 N(\varepsilon_F ) \Gamma
\over 3 \pi \Delta_0 } \left( 1 + {3 \pi^2 T^2\over 4
\Delta_0^2} \right),
\label{c_time}
\end{eqnarray}
where $v_F$ is the Fermi velocity of electrons in the ab
planes.  Both $\Lambda_{ab}$ and $\Lambda_c$ are finite at
$T=0$.  This is a direct consequence of the finite density
of states at the Fermi level in a disordered $d$-wave
superconductor.

From the results for previously obtained $K_\mu$ and Eqs.
(\ref{ab_time}) and (\ref{c_time}), it is straightforward to
calculate the normalized superfluid density.  In the limit
$T\ll \Gamma$, we have
\begin{eqnarray}
{\rho_s^{ab} (T)\over \rho_s^{ab} (0) } &\sim& 1 - \alpha_{ab} 
{\Delta_0 \over \Gamma} \left( {T \over \Delta_0}\right)^2 ,
\\
{\rho_s^c(T)\over \rho_s^c (0) } &\sim& 1 - \alpha_c 
{8 \pi\over 3 } {\Gamma\over \Delta_0} \left( 
{T\over \Delta_0}\right)^2 ,
\end{eqnarray}
where $\alpha_{ab}$ and $\alpha_c$ are two system dependent
dimensionless constants of order unity.  Both $\rho_s^{ab} (T) /
\rho_s^{ab} (0)$ and $\rho_s^c(T)/ \rho_s^c (0) $ now
vary quadratically with $T$, but the coefficients of the
$T^2$ terms are very different for the $ab$ plane and
$c$-axis responses.  The $T^2$ term of the $c$-axis response
is a factor $(\Gamma / \Delta_0)^2$ weaker than that in
the $ab$ plane due to the $(\cos k_x - \cos k_y)^2$ factor
in $\varepsilon_{\perp k}$.  When $\Gamma \ll T \ll
\Delta_0$, the finite lifetime effect is not important, the
intrinsic behaviors of $\rho_s^{ab} (T)$ and $\rho_s^c (T)$,
Eqs.  (\ref{sab_ani}) and (\ref{T_five}), should be
recovered.  The crossover temperature from the disordered
$T^2$ behavior to the intrinsic $T$ or $T^5$ behavior is
given by the scattering rate $\Gamma$.

\subsection{Impurity assisted hopping}

In Sec. \ref{sec3}, we showed that $\varepsilon_{\perp k}
\propto (\cos k_x - \cos k_y)^2$ for tetragonal HTSC.
Effectively, we can represent this interlayer hopping
integral by the following Hamiltonian:
\begin{equation}
H_\perp = \sum_{i\alpha\alpha^\prime} t_\perp D_{\alpha}
D_{\alpha^\prime} c^\dagger_{i+\alpha + {\hat z}} c_{i + 
\alpha^\prime } + h.c , \label{int_layer}
\end{equation}
where $D_\alpha$ is a function of $\alpha$ with
$d_{x^2-y^2}$ symmetry:  $D_\alpha = 1/2$ or $-1/2$ if
$\alpha = \pm {\hat x}$ or $\pm {\hat y}$, respectively.
This Hamiltonian is clearly valid only for a clean system.
If, however, impurities within or between two layers block or
modify the interlayer hopping at some of the sites, then a
scattering potential dependent interlayer hopping
Hamiltonian
\begin{equation} 
H_{imp}^{(1)} =
\sum_{i\alpha\alpha^\prime} W_i D_{\alpha} D_{\alpha^\prime}
c^\dagger_{i+\alpha + {\hat z}} c_{i + \alpha^\prime } + h.c
\end{equation} 
should be added to Eq.  (\ref{int_layer}), where $W_i$ is
the random scattering potential.

In $H_{imp}^{(1)} $, we have implicitly assumed that the
scattering potential preserves locally the $d_{x^2-y^2}$
symmetry of the interlayer hopping integral.  In real
materials, however, this local symmetry may not always be preserved
by random scatterers.  The contribution from
those random scatterers which do not preserve this local
symmetry should also be included in the impurity assisted
hopping integral.  For simplicity, we assume that the
interlayer hopping assisted by those scatterers has the form
\begin{equation}
H_{imp}^{(2)} =  \sum_{i}V_i ( c_{i+{\hat z}}^\dagger c_i +
h.c.  ) ,
\end{equation}
where $V_i$ is the scattering potential.  Therefore, the
total impurity assisted hopping integral of electrons is
$H_{imp} = H_{imp}^{(1)} + H_{imp}^{(2)}$.

In HTSC, the interlayer coupling is weak.  Thus we can take
$H_{imp}$ as a perturbation to calculate its contribution to
$\rho_s^c$.  Up to the second order approximation in
$H_{imp}$, we find that
\begin{eqnarray}
 \rho_{s,imp}^c &\sim & -{4\over \beta}\sum_{k, k^\prime}
 \langle M_{k,k^\prime}  M_{k^\prime ,k} \rangle_{imp} 
 \int_{-\infty}^\infty {\rm d}\omega f(\omega) 
\nonumber \\
  & & {\rm Tr}
 \left[ {\rm Re} G(k , \omega ) {\rm Im} G(k^\prime , 
 \omega ) \right.
\nonumber  \\
& & \left. - {\rm Re} G(k , \omega ) \tau_3 {\rm Im} G(k^\prime ,  
 \omega )  \tau_3 \right],  
\end{eqnarray}
where $M_{k,k^\prime} = V_{k-k^\prime} + W_{k-k^\prime}
\gamma_k \gamma_{k^\prime}$ and $\gamma_k = (\cos k_x - \cos
k_y)$.  The impurity average of the assisted hopping matrix
element $\langle M_{k, k^\prime} M_{k^\prime ,
k}\rangle_{imp} $ can be separated into three terms
$$
 \langle M_{k,k^\prime}  M_{k^\prime ,k} \rangle_{imp}
= V_{k-k^\prime}^{(1)} + V_{k-k^\prime}^{(2)} \gamma_k 
\gamma_{k^\prime} + V_{k-k^\prime}^{(3)} \gamma_k^2 
\gamma_{k^\prime}^2,
$$
where $V^{(1)}_k = \langle V_{k} V_{-k} \rangle_{imp}$,
$V^{(2)}_k = \langle V_{k} W_{-k} + W_{k} V_{-k}
\rangle_{imp}$, $V^{(3)}_k = \langle W_{k} W_{-k}
\rangle_{imp}$.  The second term has the same form as 
Hirschfeld {\it et al} \cite{Hirs97} assumed for the
scattering potential with pair fluctuations between layers.
But here this term comes purely from the single particle
scattering potential.

The contribution of the first term of $\langle
M_{k,k^\prime} M_{k^\prime ,k} \rangle_{imp}$ to
$\rho_{s,imp}^c$ was first considered by Radtke {\it et
al} \cite{Radtke96}.  If $V^{(1)}_{k -k^\prime}$ is
independent on $k-k^\prime$ (this corresponds to a diffuse
transmission of electrons between layers), then the
contribution of this term to $\rho_s^c$ is zero.  This is a
special consequence of an isotropic scattering matrix
element and does not hold when anisotropy due to the
$d$-wave superconducting fluctuation is included.  If,
however, the scattering potential is anisotropic and long
range correlated, for example if $V_{k-k^\prime}^{(1)}$ has
the Lorentzian form \cite{Radtke96} $V_{k-k^\prime}^{(1)} =
V_1 k_F\delta k / [ ({\bf k}-{\bf k}^\prime )^2 + (\delta
k)^2]$, then in the strong forward scattering limit $\delta
k / k_F \rightarrow 0$, $\rho^c_{s,imp}$ behaves as $T^2$ at
low $T$ \cite{Xiang96b}:
\begin{equation} 
 \rho^{c,1}_{s,imp} \approx
 2 \pi V_1 \Delta_0 N^2(\varepsilon_F) \left[ 1 -8 \ln 2 \left( {T\over
 \Delta_0}\right)^2 \right] .
\end{equation}
For a small but finite $\delta k$, $\rho_{s,imp}^c $ behaves
similarly to the $\delta k \rightarrow 0$ case, but the
overall amplitude of $\rho^c_{s,imp}$ decreases with
increasing $\delta k$.

The contribution of the second term of $\langle M_{k,
k^\prime} M_{k^\prime ,k}\rangle $ to $\rho_s^{c,imp}$ has
been studied recently by Hirschfeld {\it et al}
\cite{Hirs97}.  If $V^{(2)}_{k -k^\prime} = V_2$ is
dependent on $k-k^\prime$, then the contribution of this
term $\rho_{s,imp}^{c}$ at low temperature is
\begin{equation}
\rho_{s,imp}^{c,2} \approx \pi V_2 N^2(\varepsilon_F)
\Delta_0 \left[ 1 - { 12\zeta (3) \over \pi } \left( 
{T\over \Delta_0} \right)^3 \right].
\end{equation}
We note that the sign of $V_2$ is very important here.  If
$V_2$ is negative, $\rho_{s, imp}^{c,2}$ is also negative,
which is obviously unphysical.  Thus when $V_2 < 0$, the 
above calculation based on the second order perturbation is
invalid and higher order corrections of $H_{imp}^{(1)}$ to
$\rho_{s,imp}^{c,2}$ should be considered.

The contribution from the third term of $\langle
M_{k,k^\prime} M_{k^\prime ,k}\rangle $ to $\rho_{s, imp}^c$
is zero if $V^{(3)}_{k-k^\prime} $ is a constant 
independent of $k-k^\prime$. The reason for this is 
that the average of $\gamma^3_k$ (one of the $\gamma_k$ is 
from the gap function) over the Fermi surface is zero.

\subsection{Summary}

The above discussion indicates that both lifetime effects
and impurity assisted hopping may give a large contribution
to $\rho_s^c$ at low $T$.  The intrinsic $T^5$ behavior of
$\rho_s^c$ is therefore observable only in samples which are
pure and of relatively low anisotropy so that the
contribution from the coherent interlayer tunneling to
$\rho_s^c$ is substantially larger than that from disorder
effects.  In all compounds and at sufficiently low $T$,
$\rho_s^c$ should eventually behave as $T^2$ or $T^3$
depending on which type of disorder effects is stronger.  If
the contribution from the lifetime effect or the first term
of $\langle M_{k,k^\prime} M_{k^\prime ,k} \rangle_{imp}$ is
larger, $\rho_s^c$ varies as $T^2$ at low $T$; whereas if
the contribution from the second term of $\langle
M_{k,k^\prime} M_{k^\prime ,k} \rangle_{imp}$ is larger,
$\rho_s^c$ varies as $T^3$ at low $T$.

Disorder has opposite effects in the $ab$ plane and along
the $c$-axis:  it weakens the $T$-dependence of the in-plane
response and strengthens the $T$-dependence of the out of
plane response.  As will be shown later, this result agrees
well with experiments.  We believe it is valid also for YBCO
related materials although our discussion in this section is
for tetragonal HTSC only .

\section{Experimental results} \label{sec5}

\subsection{Technique}

There are four main techniques for measuring the penetration
depth $\lambda (T)$, namely optical conductivity, microwave
surface impedance, muon spin relaxation, and low field ac
magnetic susceptibility.  The optical conductivity technique
has been used to determine the absolute value of $\lambda
(0)$; measurements as a function of temperature are in
principle possible but time consuming.  The microwave method
can be used to determine the relative change of $\lambda
(T)$ with respective to $\lambda (0)$ accurately, but not
$\lambda (0)$ itself.  Muon spin relaxation measurements can
be used to determine the in-plane penetration depth at
$T=0$.  The method we used is the ac susceptibility
technique, with which we can simultaneously determine
$\lambda (0)$ directly as well as $\lambda (T)$ both
parallel and perpendicular to the CuO planes.

Two susceptometers were used in our ac susceptibility
measurements.  One susceptometer is home-made and is used to
measure the susceptibility in the $T$ range 1.2 - 40 K.  The
second susceptometer, which is commercially available, was
used to obtain data in the temperature range 4.2 - 150 K.
To achieve the maximum sensitivity, powder samples, with
grain sizes typically of the order of a few microns, were
used.  To align the samples, the high-$T_c$ powders were
mixed with a 5min fast curing epoxy and placed in a high
static field at room temperature until the epoxy cures.  The
misalignment of the CuO planes of the samples is generally
of the order of 1$^\circ$ for more than 95\% of the grains.
However, the surface of the sample can be easily
contaminated when the bulk polycrystalline sample is ground
into fine powder.  To obtain a clean surface, we prepared
the sample either by grinding the bulk ceramic in an argon
atmosphere, or by heat treating the powder after grinding in
air \cite{Pana97d,Pana97a,Pana96a,Pana96b,Pana97c,Chrosch96}

If we assume that all grains are spherical, then $\lambda$
can be estimated from the magnetization according to the
formula  \cite{Porch93} 
\begin{equation} 
{M\over M_0} ={\int {\rm d} R
\left( 1 - {\displaystyle 3\lambda \over \displaystyle R}
\coth {\displaystyle R\over \displaystyle \lambda} +
{\displaystyle 3 \lambda^2 \over \displaystyle R^2 } \right)R^3 g(R)
\over \int {\rm d} R R^3 g(R) } \label{lambda}
\end{equation}
where $R$ is the grain radius and $g(R)$ is the distribution
function of $R$ \cite{Cooper88}.  $M_0$ is the diamagnetic
moment of a unpenetrated sample and it can be calculated
from the mass of the powder.

To determine the grain size distribution function $g(R)$, we
took scanning electron microscopy photographs for the
powders and measured the dimensions of all grains along two
perpendicular axes.  These photographs also show that more
than 80\% of the grains are nearly spherical.  The
distribution function $g(R)$ such determined may have a
certain error.  We find that the value of $\lambda$ obtained
from Eq.  (\ref{lambda}) is not sensitive to the detailed
form of $g(R)$.  For example, reducing the number of grains
measured by half leads to only a 3\% change in $\lambda$ and
no change in the $T$ dependence of $\lambda$.  Deviation
from the spherical geometry tends to increase the actual
surface-area to volume ratio, and effectively change the
distribution function $g(R)$.  This effect is less important
when $\lambda \sim R$ where the ratio of the penetrated to
unpenetrated sample volume is similar for both spherical and
non-spherical specimens.  A large grain contributes a larger
diamagnetic signal than a small one, thus the measured
susceptibility is very sensitive to the number of large
grains when the majority of grains are comparable to the
penetration depth in size.  The uncertainty in the grain
size can be reduced by sieving or sedimenting the powder
before alignment.

Intergrain contacts may affect the behavior of $\lambda$
as well.  However, from the analysis of the linearity of the
susceptibility in an ac field from 0.3 to 10 Gauss with
frequency from 33 to 667 Hz, we find that the error resulted
from this effect is less than 2\%.  Another possible source
of error comes from the error in $M_0$.  To determine
$M_0$ accurately, it is important to estimate accurately the
mass of superconductor in the superconductor-epoxy
composite.  The error in $M_0$ is mainly from the
sedimentation of the grains during the alignment procedure.
To reduce this error we used a fast curing epoxy and ensured
that our results were reproducible between different samples
cut from the same
superconductor-epoxy composite.  We note that none of the
above errors affect the temperature dependence of $\lambda$.
A uncertainty of 5\% in the degree of grain alignment (which
is higher than we usually encounter) yields at most 25\% and
8\% error in $\lambda_{ab}(0)$ and $\lambda_c (0)$,
respectively.  The corresponding uncertainty in the low
temperature dependence in $\lambda_{ab}(T)/ \lambda_{ab}(0)$
is at most 10\% whereas it is negligible in $\lambda_c (T) /
\lambda_c (0)$.

When the external field is parallel to the c-axis of the
grains, the value of $\lambda$ calculated from Eq.
(\ref{lambda}) is simply the in-plane penetration depth
$\lambda_{ab}$ since the diamagnetic screening current flows
entirely within the ab-plane.  However, when the external
field is parallel to the ab-plane, the screening currents
flow in both the ab plane and c axis.  In this case, only an
effective penetration depth, $\lambda_{eff}$, can be
obtained from Eq.  (\ref{lambda}).  An approximate solution
\cite{Pana96b,Porch93}, $\lambda_{eff} \sim 0.7\lambda_c$ in
the limit $\lambda_c\gg \lambda_{ab}$ can be used to
estimate $\lambda_c$.  The systematic error in $\lambda_c$
obtained in such a way is less than $\pm$2.5\% even when
$\lambda_c(0) / \lambda_{ab} \sim 4$.

\subsection{Results}

\begin{table}
\caption{$T_c$ and zero temperature $\lambda_{ab}(0)$ and 
$\lambda_c(0)$ for several high-$T_c$ materials.}
\begin{tabular}{lcccc}

\hline
Compounds & $T_c$ (K) & $\lambda_{ab}(0)$ (\AA) & 
 $\lambda_{c}(0)$ (\AA)  & References \\
 \hline

${\rm Ba_{0.6}K_{0.4}BiO_3}$ & 25 & \multicolumn{2}{c}{ 3500 } &  
 \cite{Pana96a} \\

${\rm HgBa_2CuO_{4+\delta}}$ & 93 & 1710 & 13600 &  
 \cite{Pana97a} \\

${\rm HgBa_2Ca_2Cu_3O_{8+\delta}}$ & 135 & 1770 & 61000 & 
 \cite{Pana96a,Pana97a}   \\

${\rm YBa_2Cu_3O_7}$ & 92 & 1400 & 12600 & 
 \cite{Pana97d,Pana96b}  \\

${\rm YBa_2Cu_3O_{6.7}}$  & 66 & 2100 & 45300 & 
 \cite{Pana97d} \\

${\rm YBa_2Cu_3O_{6.57}}$  & 56 & 2900 & 71700 & 
 \cite{Pana97d} \\

${\rm YBa_2(Cu_{1-x}Zn_x)_3O_7}$ (x = 0.02) & 68 & 2600 & 14200 & 
 \cite{Pana96b}   \\

 \hspace{3.25cm} (x = 0.03) & 55 & 3000 & 15500 & 
  \cite{Pana96b}   \\

 \hspace{3.25cm}  (x = 0.05) & 46 & 3700 & 16400 & 
  \cite{Pana96b}   
\\ \hline

\end{tabular}
\end{table}

We have measured the penetration depth of slightly overdoped
${\rm HgBa_2CuO_{4 +\delta}}$ (Hg1201), slightly underdoped
${\rm HgBa_2Ca_2Cu_3O_{8 +\delta} }$ (Hg1223), ${\rm
YBa_2Cu_3O_{7 -\delta} }$ ($\delta$ = 0.0, 0.30 and 0.43)
and ${\rm YBa_2(Cu_{1-x} Zn_x)_3O_7 }$ (x=0.02, 0.03 and
0.05).  A few parameters for these materials are given in
Table I.  $\gamma = \lambda_c (0) / \lambda_{ab} (0) $ is an
important quantity in characterizing the interlayer
coupling.  The one-layer tetragonal compound Hg1201 has the
lowest anisotropy among all tetragonal HTSC, $\gamma \sim
8$.  The three-layer tetragonal compound Hg1223 has a higher
$T_c$ (135K) as well as a higher $\gamma$ ($\sim 34$) than
Hg1201.  The anisotropic ratios of YBCO materials depend
strongly on the doping level.  With increasing oxygen
deficiency, $T_c$ of YBCO drops but $\gamma$ increases.
$\gamma$ for YBCO$_7$, YBCO$_{6.7}$, YBCO$_{6.57}$ are 9,
22, and 26, respectively.  Zn substitution replaces the
planar Cu atoms and strongly suppresses $T_c$.  In Zn doped
YBCO, both $\lambda_{ab}$ and $\lambda_c$ were found to
increase with Zn concentration, but $\gamma$ drops from 9
for $x=0$ to 4.4 for $x=0.05$.

\begin{figure} 
\begin{center}
\leavevmode\epsfxsize=8cm
\epsfbox{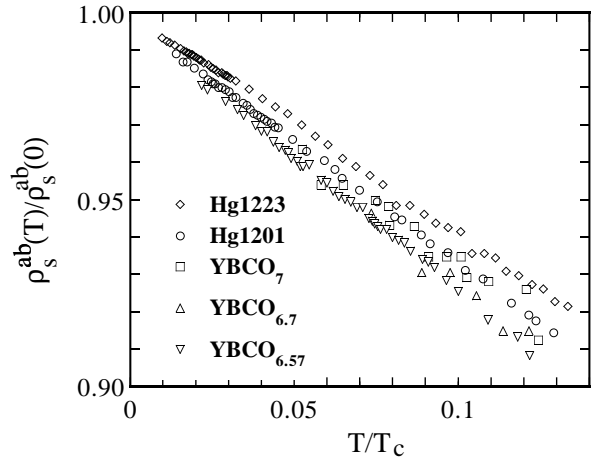}
\caption{Normalized in-plane superfluid density
$\rho_s^{ab}(T) / \rho_s^{ab}(0)$ vs $T/T_c$ for Hg1201,
Hg1223, YBCO$_7$, YBCO$_{6.7}$, and YBCO$_{6.57}$ 
at low temperature.  }
\label{chris1} 
\end{center}
\end{figure}

Fig.  \ref{chris1} shows the reduced in-plane superfluid
density $\rho_s^{ab}(T) / \rho_s^{ab}(0)$ as a function of
the reduced temperature $T/T_c$ for the three YBCO and two
Hg-related compounds at low $T$.  The linear $T$ dependences
of $\rho_s^{ab}$ at low $T$ in these materials are
consistent with the recent microwave measurement results of
YBCO \cite{Bonn95}, ${\rm Bi_2Sr_2CaCu_2O_{8+\delta}
}$ \cite{Jacobs95,Lee96} and ${\rm Tl_2Ba_2CuO_{6+\delta} }$
 \cite{Broun97} single crystals, indicating that the
high-$T_c$ pairing has indeed $d$ symmetry, independent on
the number of CuO$_2$ planes per unit cell, carrier
concentration, crystal symmetry, anisotropy and the presence
of chains.

For all the compounds, both the zero temperature penetration
depth and the the slope of the linear $T$ term of
$\lambda_{ab}$ are very different.  For example, the
coefficients of the linear $T$ term of $\lambda_{ab}$ are
6.5$\AA / K$, 4.2$\AA / K$, 4.8$\AA / K$, 12$\AA / K$, and
20$\AA / K$ for Hg1201, Hg1223, YBCO$_7$, YBCO$_{6.7}$, and
YBCO$_{6.57}$, respectively.  We note, however, that the
reduced superfluid density $\rho_s^{ab}(T) / \rho_s^{ab}(0)$
as a function of $T/T_c$ does not change much for all five
samples.  If we fit the data with the weak-coupling BCS
result, Eq.  (\ref{sab_ani}), we find that $\Delta_0$ scales
approximately with $T_c$, $\Delta_0 \sim 2 T_c$, for these
materials.

This approximate scaling behavior of $\Delta_0$ with $T_c$
is undoubtedly an important property of HTSC.  However,
AREPS \cite{Rand97} and tunneling \cite{Renner98}
experiments have shown that the maximum energy gap is much
larger than what expected from the weak-coupling BCS theory
and $\Delta_0 / T_c$ increases rapidly with decreasing
doping in the underdoped regime.  A possible explanation for
this discrepancy is that the gap amplitude $\Delta_0$ in the
$d$-wave gap function $\Delta_k = \Delta_0 \cos 2 \phi$ is
strongly $\phi$-dependent rather than simply a constant:
the energy gap given by ARPES or tunneling spectroscopy
measurements is the maximum gap, namely $\Delta_{max}=
\Delta_0(\phi)$ at $\phi =0$ or $\pi /2$; whereas $\Delta_0$
determined from low $T$ in-plane penetration depth
measurement is that around the gap nodes, namely at $\phi =
\pi /4$.

This explanation is in fact consistent with recent ARPES
measurements \cite{Norman97} which show that in the normal
phase of underdoped materials, a pseudogap opens first
around $\phi =0$, $\pi /2$ at high temperature and then
spreads towards the d-wave gap nodes with decreasing $T$.
Thus the effect of the pseudogap on the superconducting gap
parameter is largest at $\phi =0$ and $\pi /2$, and smallest
at $\phi = \pi /4$.  Hence, although $\Delta_0$ at $\phi
=0$, $\pi /2$ does not scale with $T_c$ in underdoped
materials, $\Delta_0$ at $\phi = \pi/4$ does.  This also
explains why in the overdoped regime, where the pseudogap is
very small if not completely absent, the superconducting gap
$\Delta_0$ obtained from ARPES and tunneling scales
approximately with $T_c$.

Along the $c$-axis, the low $T$ behavior of the superfluid
response function is strongly sample dependent.  A common
feature observed in all the materials we measured is
that the variation of $\rho_s^c(T) /\rho_s^c(0)$ with $T$ is
weaker than its in-plane counterpart.

\begin{figure} 
\begin{center}
\leavevmode\epsfxsize=8cm
\epsfbox{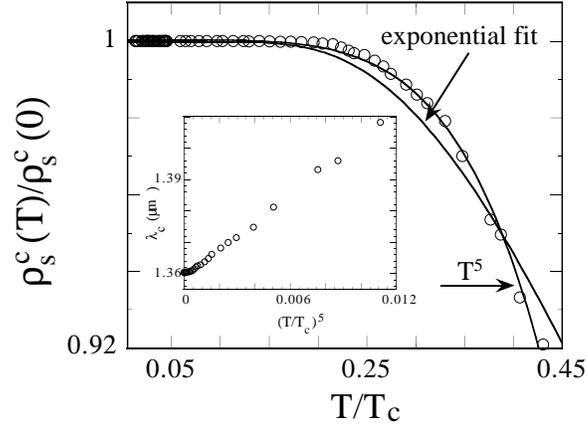}
\caption{$\rho_s^c(T) / \rho_s^c(0)$ vs $T/T_c$ for Hg1201
at low $T$.  A comparison between a $T^5$ power law
and an exponential fit to the experimental data is given.
The inset shows the experimental data of the $c$-axis
penetration depth $\lambda_c$ as a function of $(T/T_c)^5$.
}
\label{chris2} 
\end{center}
\end{figure}

For Hg1201 (Figure \ref{chris2}), we find that $\rho_s^c$
behaves as $T^5$ at low $T$ in agreement with theoretical
prediction Eq.  (\ref{T_five}).  From the fitting, we find
that $\Delta_0 /T_c \sim 2.37$ which is close to the weak
coupling BCS value, $\Delta_0 /T_c \sim 2.14$, for a
$d$-wave superconductor.  In Figure \ref{chris2}, a
comparison between the $T^5$ and an exponential fit to the
experimental data is also shown.  It is clear that the
exponential fit is inferior to the $T^5$ power law fit.

The agreement between the theoretical prediction and the
experimental result for the $T^5$ behavior of $\rho_s^c$
shows unambiguously that the gap nodes are located along the
zone diagonals and the high-$T_c$ pairing has indeed the
$d_{x^2-y^2}$ symmetry.  Thus the c-axis penetration depth
measurement can reveal not only the existence of the gap
nodes (as in the in-plane penetration depth measurement),
but also the positions of the nodes on Fermi surface.  This
agreement also shows that the Cu $4s$ orbital is indeed very
important for understanding the interlayer dynamics and that
the low energy excitations of HTSC are governed by the
strongly hybridized Cu $d_{x^2-y^2}$ and bonding O $2p$ band
(if other bands had contribution to $\rho_s^c$, $\rho_s^c$
would deviate noticeably from $T^5$ at low $T$ since the
effective $c$-axis hopping integrals for these bands do not
have the property $\varepsilon_\perp \propto (\cos k_x -
\cos k_y)^2$).

\begin{figure} 
\begin{center}
\leavevmode\epsfxsize=8cm
\epsfbox{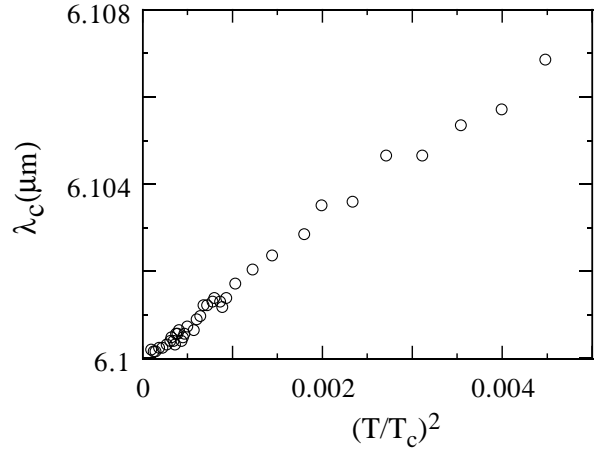}
\caption{Low $T$ $\lambda_c$ vs $(T/T_c)^2$ for Hg1223.  }
\label{chris3} 
\end{center}
\end{figure}

The $T^5$ behavior of $\rho_s^c$ for tetragonal HTSC is very
small at low $T$.  This behavior can be observed in samples
which are pure and of low anisotropy so that the coherent
inter-layer hopping of electrons is the main contribution to
$\rho_s^c$.  In highly anisotropic materials, however, the
coherent tunneling component of the supercurrent along the
$c$-axis is significantly reduced and the contribution from
incoherent tunneling induced by disorder scattering may
dominate the low $T$ behavior of $\rho_s^c$.  Thus   the
low $T$ behavior of $\rho_s^c$ for a highly anisotropic HTSC is
expected to be different than for HTSCs with lower anisotropy.

For Hg1223, the $c$-axis penetration depth $\lambda_c (T)$
varies quadratically with $T$ at low $T$ (Fig.
\ref{chris3}).  This $T^2$ term probably arises from
disorder effects which, as discussed in Sec.  \ref{sec4},
can give rise to a $T^2$ term in $\rho_s^c (T)$ without
significantly affecting the linear $T$ term in
$\lambda_{ab}(T)$.

\begin{figure} 
\begin{center}
\leavevmode\epsfxsize=8cm
\epsfbox{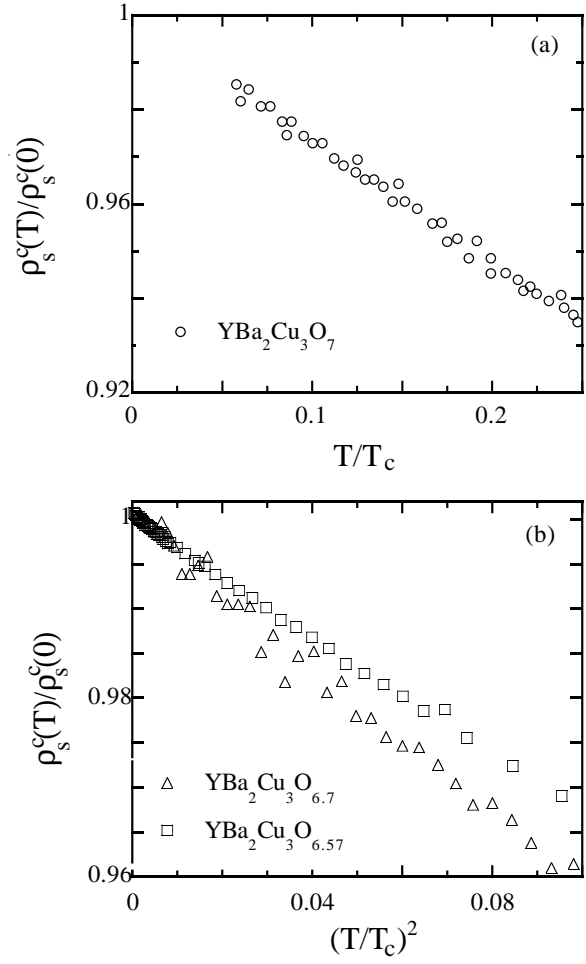}
\caption{$\rho_s^c(T) / \rho_s^c(0)$ as a function of (a)
$T/T_c$ for YBCO$_7$ and (b) $(T/T_c)^2$ for YBCO$_{6.7}$
and YBCO$_{6.57}$ at low $T$.  }
\label{chris4} 
\end{center}
\end{figure}

Figures \ref{chris4}(a) and \ref{chris4}(b) shows the
normalized $c$-axis superfluid densities as functions of
$T/T_c$ for YBCO$_7$ and $(T/T_c)^2$ for YBCO$_{6.7}$ and
YBCO$_{6.57}$, respectively.  $\lambda_c$ for YBCO$_7$
($\gamma = 9$) exhibits a linear $T$ dependence at low $T$
but the relative change is about a factor of two smaller
than in $\lambda_{ab}(T)$ / $\lambda_{ab}(0)$.  As discussed
earlier this probably arises from the effects of hopping
between planar and chain bands present in YBCO (see Eq.
(\ref{ybco_rho})).  By removing oxygen from YBCO$_7$ one can
reduce the effects of the chains.  $\lambda_c (T)$ of
YBCO$_{6.7}$ ($\gamma\sim 22$) and YBCO$_{6.57}$ ($\gamma
\sim 25$) vary as $T^2$, like Hg-1223 which has similar
anisotropy.

\begin{figure} 
\begin{center}
\leavevmode\epsfxsize=8cm
\epsfbox{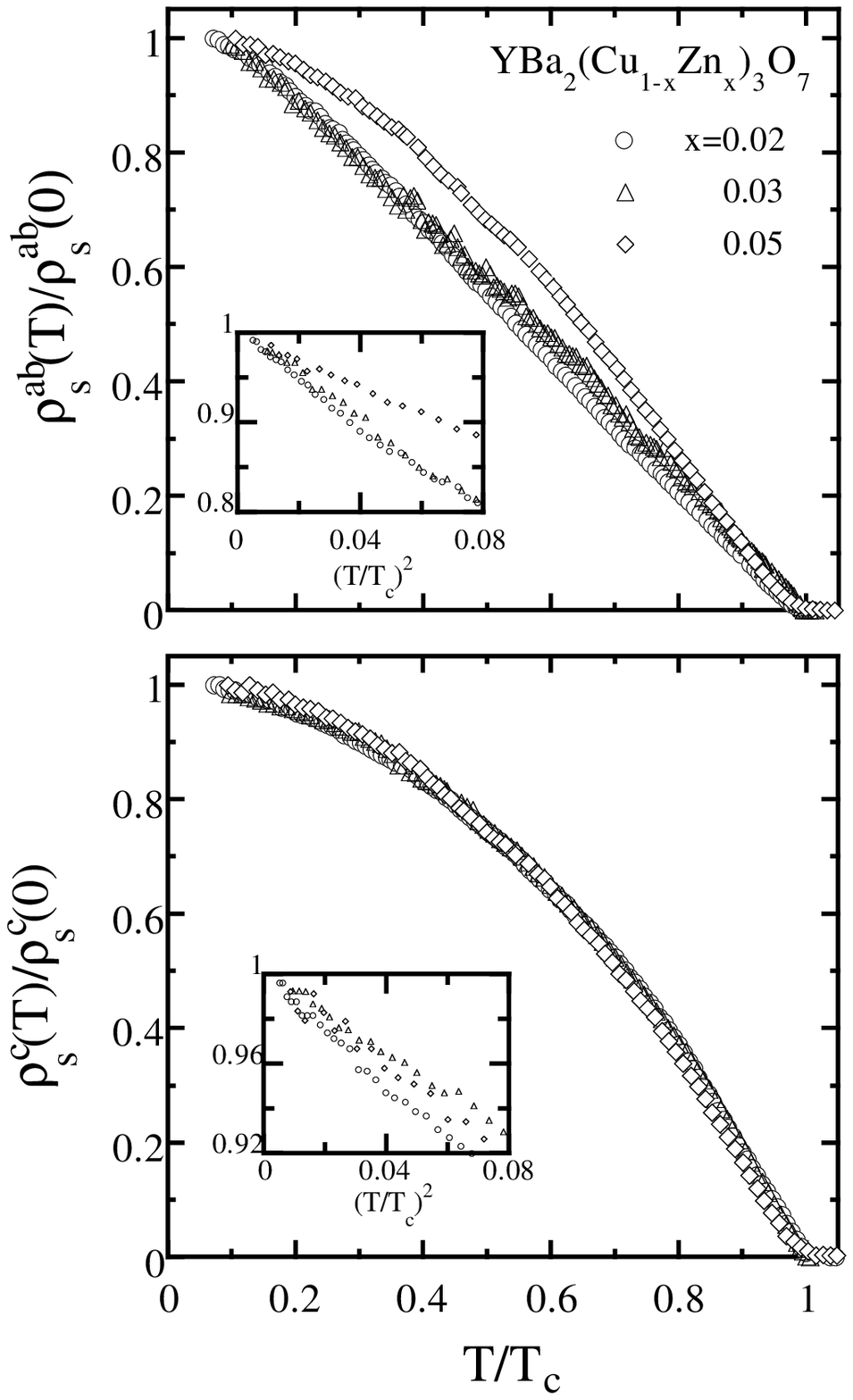}
\caption{Normalized in-plane (a) and $c$-axis (b) superfluid
density vs $T/T_c$ for ${\rm YBa_2(Cu_{1-x}Zn_x)_3O_7}$ with
x=0.02, 0.03 and 0.05.  Insets show the corresponding
reduced superfluid tensor as a function of $(T/T_c)^2$ at
low T.  }
\label{chris5} 
\end{center}
\end{figure}

\begin{figure} 
\begin{center}
\leavevmode\epsfxsize=8cm
\epsfbox{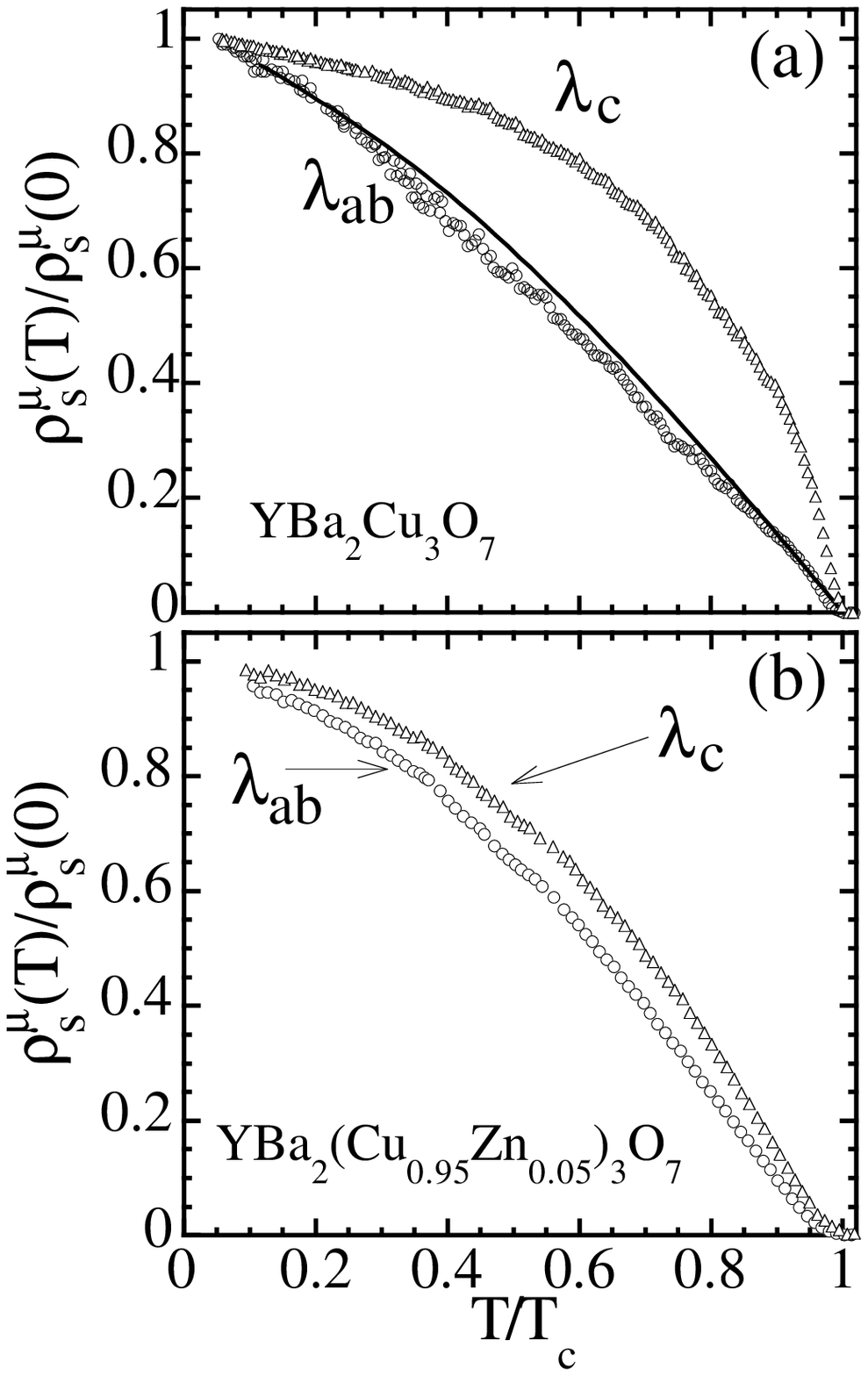}
\caption{Comparison between the reduce in-plane 
and out-of-plane superfluid density as a function of 
$T/T_c$ for (a) ${\rm YBa_2Cu_3O_7}$ and (b) ${\rm 
YBa_2(Cu_{0.95}Zn_{0.05})_3O_7}$. The solid line 
in (a) is the weak-coupling BCS result for a $d$-wave 
superconductor.  }
\label{chris6} 
\end{center}
\end{figure}

Figure \ref{chris5} shows the normalized in-plane (a) and
$c$-axis (b) superfluid density as a function of $T/T_c$ for
${\rm YBa_2(Cu_{1-x}Zn_x)_3O_7}$ with x=0.02, 0.03 and 0.05.
As shown in Table I, $\lambda_{ab}(0)$ increases very
quickly with Zn doping.  This fast increase of
$\lambda_{ab}(0)$ is consistent with an increase in the
residual density of states as revealed by the measurement
of low temperature specific heat \cite{Loram}.  The increase of $\lambda_c
(0)$ with Zn concentration is relatively slower, which may
be due to an enhanced c-axis coupling caused by Zn doping.
The low T dependences of both $\lambda_{ab}$ and $\lambda_c$
change from $T$ to $T^2$ with Zn doping due to impurity
scattering.  Another interesting property of Zn doped YBCO
is that with increasing doping both $\lambda_{ab}(0)$ /
$\lambda_{ab}(T)$ and $\lambda_c (0)$ / $\lambda_c (T)$
gradually approach each other, and for x = 0.05 they have
almost the same $T$ behavior (Fig.  \ref{chris6}).

\section{Summary} \label{sec6}

In summary, the low $T$ behavior of the superfluid tensor
has been systematically investigated.  We found that
 the low $T$ slope of $\rho_s^{ab}(0) /
\rho_s^{ab}(T)$ scales approximately with $T_c$,
irrespective of their doping level, anisotropy and chemical
structures.  If we assume that the weak coupling BCS theory
is applicable in the low $T$ regime of HTSC, this result
implies that the gap amplitude $\Delta_0$ scales
approximately with $T_c$ for all high-$T_c$ materials.

For tetragonal HTSC, we gave a simple symmetry argument
which shows that the $c$-axis hopping matrix element
$\varepsilon_\perp (k_\parallel )\propto (\cos k_x -\cos
k_y)^2$ and that for low anisotropic Hg1201 $\rho_s^c$
behaves as $T^5$ at low $T$.  For more anisotropic
materials, such as Hg1223, $\rho_s^c (T)$ behaves as $T^2$
at low $T$ due to disorder effects.

The planar anisotropy of electromagnetic response of clean
YBCO-type structures gives a useful probe of the microscopic
state of cuprate superconductors.  If the superfluid density
on the CuO chains is induced purely by the proximity effect,
$\lambda_b$ is shown to behave as $\sqrt{T}$ at low $T$.  If
on the other hand, the pair tunneling (or Josephson
tunneling) process between CuO chains and CuO planes is
important, $\lambda_b$ varies linearly with $T$ at low $T$.
The experimental data of ${\rm YBa_2Cu_3O_{7-\delta}}$ agree
with the pair tunneling picture, but the low $T$ behavior of
${\rm YBa_2Cu_4O_8}$ seem to be consistent with the result
of the proximity model.  Why the low $T$ $\lambda_b$ of
${\rm YBa_2Cu_3O_{7-\delta}}$ and that of ${\rm
YBa_2Cu_4O_8}$ behave so differently requires further
theoretical investigation.

The effect of disorder on the low $T$ behavior of the
superfluid tensor is more apparent in heavily underdoped or
Zn doped YBCO.  In YBCO$_{6.57}$ and YBCO$_{6.7}$,
$\rho_s^{ab}$ and $\rho_s^c$ vary linearly and quadratically
with $T$ at low $T$, respectively.  For Zn-doped YBCO, both
$\rho_s^{ab}$ and $\rho_s^{c}$ behave as $T^2$ at low $T$.

\section*{References}


\eject

\end{document}